\documentclass[a4paper,twoside]{article}
\usepackage{epsfig}
\usepackage{subcaption}
\usepackage{calc}
\usepackage{amssymb}
\usepackage{amstext}
\usepackage{amsmath}
\usepackage{amsthm}
\usepackage{multicol}
\usepackage{pslatex}
\usepackage{natbib}
\let\bibhang\relax
\usepackage{apalike}
\usepackage{algorithm2e}
\usepackage[bottom]{footmisc}
\usepackage{SCITEPRESS} 
\usepackage{dirtytalk}
\usepackage{float}
\usepackage{etoolbox}
\usepackage{xcolor}

\newbool{double_blind}
\booltrue{double_blind} 
\boolfalse{double_blind}

\renewcommand{\cite}{\citep}

\newcommand{\db}[2]{\ifbool{double_blind}{#1}{#2}}

\newcommand{\AISPRINT}{\db{[EU Project]}{AI-SPRINT}}
\newcommand{\AISPRINTName}{\db{[EU Project Full Title]}{Artificial Intelligence in Secure PRIvacy-preserving computing coNTinuum}}
\newcommand{\POPNAS}{\db{[CustomNAS]}{POPNAS}}
\newcommand{\PYCOMPS}{\db{[DistributionLib]}{PyCOMPS/dislib}}
\newcommand{\SGDE}{\db{[GenerativeFed]}{SGDE}}
\newcommand{\PPC}{\db{[PrivacyModel]}{Privacy Preserving Component}}
\newcommand{\SAIR}{\db{[RuntimeProf]}{SPACE4AI-R}}
\newcommand{\SAID}{\db{[DesignProf]}{SPACE4AI-D}}
\newcommand{\IM}{\db{[InfrastructureManager]}{IM}}
\newcommand{\OSCAR}{\db{[EnvironmentHandler]}{OSCAR}}
\newcommand{\SCONE}{\db{[SecurityTool]}{SCONE}}
\newcommand{\AMS}{\db{[MonitoringSystem]}{AMS}}
\newcommand{\BSC}{\db{[Author]}{BSC}}

\begin{document}

% \title{Harnessing the Computing Continuum: The AI-SPRINT Project's Impact across Personalized Healthcare, Maintenance and Inspection, and Farming 4.0}
\title{Harnessing the Computing Continuum across Personalized Healthcare, Maintenance and Inspection, and Farming 4.0}

\ifbool{double_blind}{%
    \author{Anonymous Authors}
}{%
    \author{\authorname{Fatemeh Baghdadi\sup{1}, Davide Cirillo\sup{1}, Daniele Lezzi\sup{2},Francesc Lordan\sup{2}, Fernando Vazquez\sup{2},
    Eugenio Lomurno\sup{3},
    Alberto Archetti\sup{3}, 
    Danilo Ardagna\sup{3},
    Matteo Matteucci\sup{3}
    }
    \affiliation{\sup{1}Department of Life Sciences, Barcelona Supercomputing Center, Barcelona, Spain}
    \email{\{fatemeh.baghdadi, davide.cirillo\}@bsc.es}
    \affiliation{\sup{2}Department of Computer Sciences, Barcelona Supercomputing Center, Barcelona, Spain}
    \email{\{daniele.lezzi, francesc.lordan, fernando.vazquez\}@bsc.es}
    \affiliation{\sup{3}Department of Electronics, Information, and Bioengineering, Politecnico di Milano, Milan, Italy}
    \email{\{eugenio.lomurno, alberto.archetti, danilo.ardagna, matteo.matteucci\}@polimi.it}
}
}

\keywords{Edge Computing, Machine Learning, Personalized Healthcare, Maintenance and Inspection, Farming 4.0}

\abstract{
The \AISPRINT{} project, launched in 2021 and funded by the European Commission, focuses on the development and implementation of AI applications across the computing continuum. This continuum ensures the coherent integration of computational resources and services from centralized data centers to edge devices, facilitating efficient and adaptive computation and application delivery. \AISPRINT{} has achieved significant scientific advances, including streamlined processes, improved efficiency, and the ability to operate in real time, as evidenced by three practical use cases. This paper provides an in-depth examination of these applications -- Personalized Healthcare, Maintenance and Inspection, and Farming 4.0 -- highlighting their practical implementation and the objectives achieved with the integration of \AISPRINT{} technologies. We analyze how the proposed toolchain effectively addresses a range of challenges and refines processes, discussing its relevance and impact in multiple domains. After a comprehensive overview of the main \AISPRINT{} tools used in these scenarios, the paper summarizes of the findings and key lessons learned. \\
\ifbool{double_blind}{%
    \textit{\textcolor{red}{The name of the EU project, the name of the tools, and the references to papers from the authors, have been anonymised for the double-blind review phase. The images have been censored to avoid references to the real tool names.}}}
    {%
}
}

\onecolumn \maketitle \normalsize \setcounter{footnote}{0} \vfill

\section{\uppercase{Introduction}}
\label{sec:introduction}

In the context of Artificial Intelligence (AI), the evolution of computing paradigms from centralized data centers to the edge of the network heralds a transformative shift in how AI applications are developed, deployed, and operated. This transition is critical to realizing the full potential of AI across a wide range of industries, from healthcare to agriculture and industrial maintenance, by leveraging the immediacy and context-aware capabilities of edge computing~\cite{article,SINGH202371}. Specifically, the edge computing paradigm is characterized by processing data directly in the devices where it is collected, such as smartphones, wearables, and IoT. Edge computing significantly reduces latency, conserves bandwidth, and enhances data privacy and security, thereby catalyzing the realization of real-time, responsive AI solutions.

\AISPRINT{} is a collective project funded by the European Commission under the Horizon 2020 program, emerging as a cornerstone in this evolving landscape. The goal of \AISPRINT{}, which stands for \AISPRINTName{}, is to provide a comprehensive design and runtime framework dedicated to accelerating the development and deployment of AI applications across the computing continuum. This continuum spans from cloud data centers to edge devices and sensors, integrating AI components to operate seamlessly and efficiently. The project's core objective is to offer a suite of tools that enable an optimal balance among application performance, energy efficiency, and AI model accuracy, all while upholding rigorous security and privacy standards. This approach addresses pivotal challenges such as flexibility, scalability, and interoperability in distributed computing environments. Moreover, \AISPRINT{} facilitates the smooth integration of security and privacy constraints from the design phase of AI applications, thus fostering resilience and trustworthiness as development guiding principles.
The aim of this work is to provide a comprehensive overview of the \AISPRINT{} framework and the results obtained over the years of its development by examining three use cases: Personalized Healthcare, Maintenance and Inspection, and Farming 4.0. These scenarios showcase the practical implications and transformative potential of the \AISPRINT{} framework.

The Personalized Healthcare use case focuses on harnessing the power of AI and wearable technologies for health monitoring. By integrating quantitative data on heart functions from wearable device sensors with qualitative lifestyle information, the project aims to develop a personalized stroke risk assessments model. This initiative is particularly critical given the prevalence of stroke among the aging population, marking it as a significant cause of death and disability globally \cite{Feigin2022}. Leveraging the \AISPRINT{} framework, the project enables efficient resource distribution and computation across the edge-to-cloud continuum, facilitating real-time, non-invasive and secure monitoring and risk assessment.

The second use case of our overview, Maintenance and Inspection, showcases the use of AI models for the early detection and analysis of windmill blade damage through the use of images captured by drones. By addressing challenges such as bandwidth limitations, unstable connections, and the drones' limited operational flight time, this use case underscores the framework's ability to enhance industrial efficiency and operator efficacy by optimizing the management of computational resources across the edge-cloud continuum. Furthermore, the integration of \AISPRINT{} tools facilitates a seamless interaction with cloud-based analytics and monitoring.

Finally, the Farming 4.0 use case innovates viticulture by harnessing the power of edge computing and AI to improve the management of phytosanitary vineyard treatments. By addressing the inefficiencies of conventional pesticide application methods, which often lead to either excessive or insufficient use \cite{Arcadia2022}, this initiative significantly mitigates environmental impact, health risks, and farmers'~economic losses.

The rest of the paper is organized as follows: Section 2 offers an overview of \AISPRINT{} and its tools to furnish context and introduce the foundational elements necessary for comprehending the remainder of the manuscript. Section 3 elaborates on the use cases, individually describing them and demonstrating the application of the tools, along with the principal outcomes achieved throughout the project's duration. Section 4 summarizes the lessons learned and consolidates the findings. Section 5 concludes the article.

\begin{figure*}[t]
\begin{center}
    \ifbool{double_blind}{
        \includegraphics[width=0.85\textwidth]{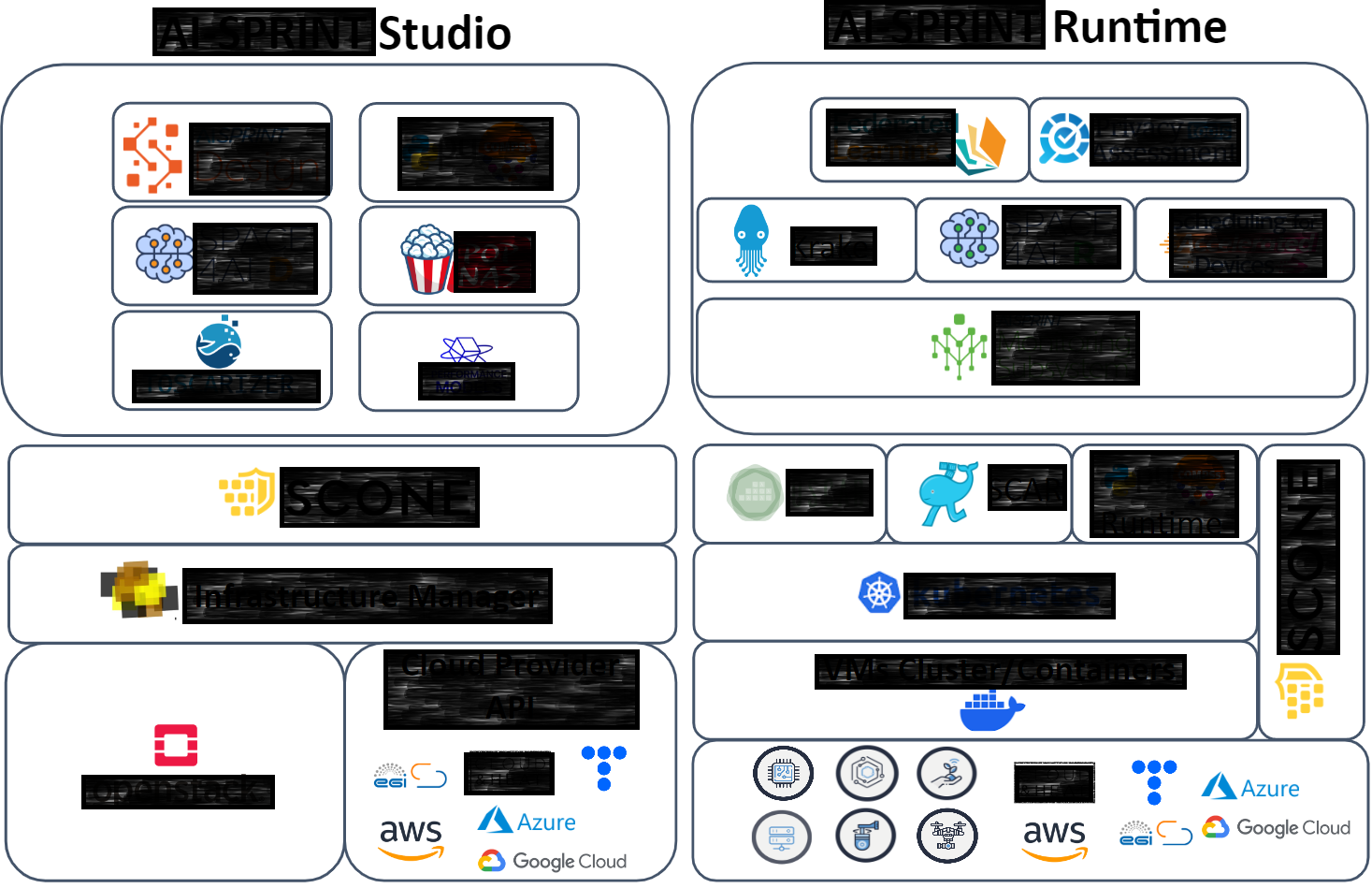}
    }{
        \includegraphics[width=0.85\textwidth]{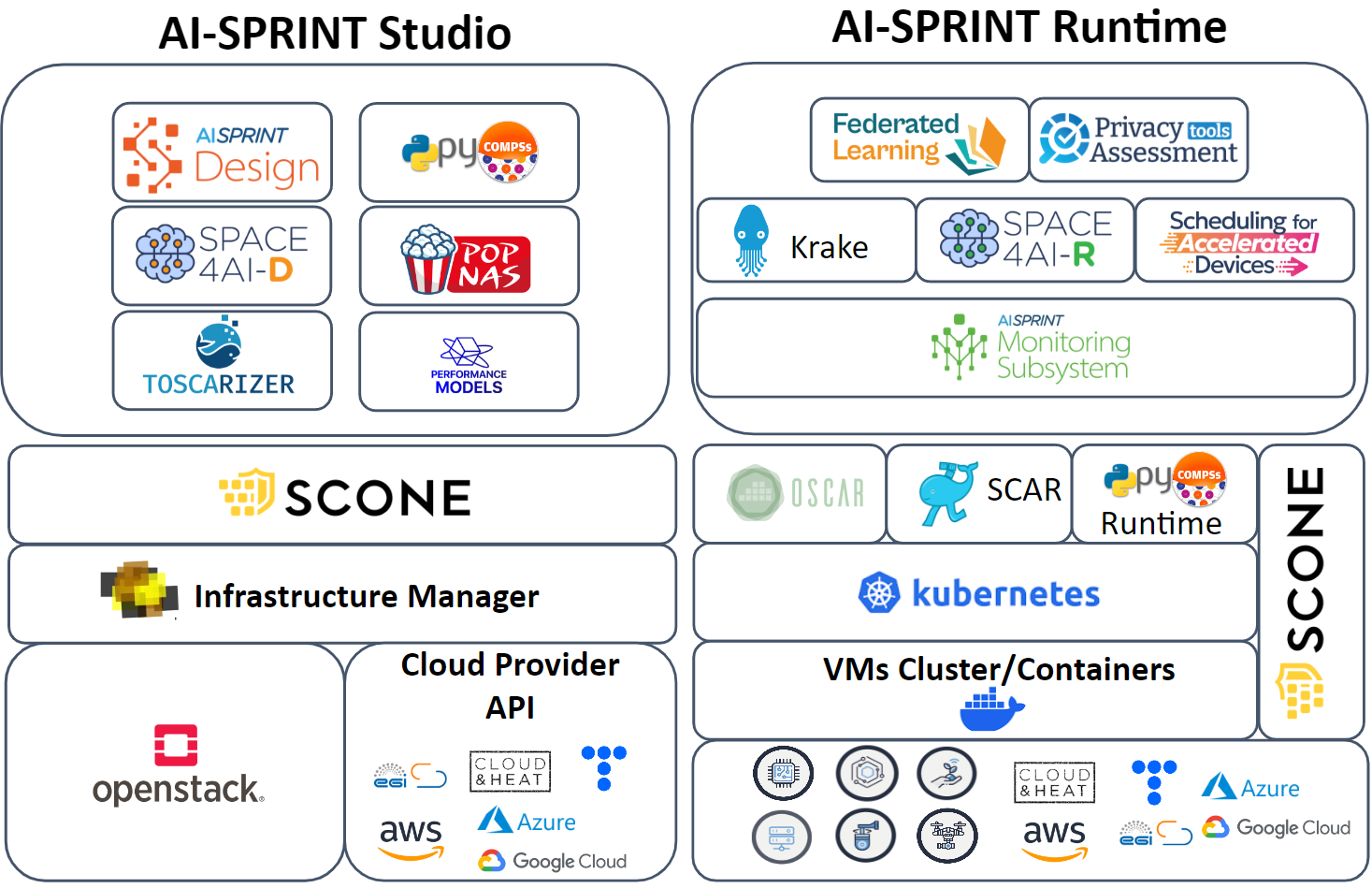}
    }
    \caption{Overview of the main \AISPRINT{} tools, divided into Studio and Runtime, and related third-party services.}
    \label{fig:tools}
\end{center}
\end{figure*}

\section{\uppercase{The Landscape of \AISPRINT{} Tools}}
The goal of the \AISPRINT{} project is to create a comprehensive framework to accelerate the development and deployment of AI applications across the spectrum of edge cloud computing. This initiative provides developers with tools to balance application performance metrics-such as latency, throughput, and energy efficiency-with AI model accuracy, all within a secure and private framework. 

In particular, the \AISPRINT{} suite provides a design environment, called \AISPRINT{} Studio, equipped with abstractions for efficient communication and parallelization using specialized resources such as GPUs and AI sensors. This environment allows performance and security requirements to be expressed through quality annotations and uses performance tools, often based on machine learning models, to automate the profiling of AI applications and the training of the most accurate models. It facilitates the deployment of deep networks in diverse computing environments and enables developers to train high-quality models with minimal machine learning (ML) expertise through AI Neural Architecture Search (NAS) algorithms. In addition, design space exploration tools help optimize resource efficiency and reduce cloud usage costs for complex applications.

On the runtime side, the \AISPRINT{} framework supports system operators with tools for seamless deployment and execution on heterogeneous architectures. Key features include support for continuous deployment, concurrent execution, resilience to node failures, automated edge-based model retraining, and the use of privacy-preserving and federated learning techniques. It also provides optimized scheduling of accelerator devices for better resource management. Packaging tools further simplify deployment by automating the generation of component placement configurations and creating application component images for different architectures, facilitating deployment across multiple hardware platforms. Deployment tools streamline resource configuration across the cloud and edge, orchestrating the deployment of AI applications without manual intervention for improved resource efficiency and scalability.

Security is built into both phases with tools for deploying compute instances in Trusted Execution Environments, securing communication channels with authentication, encryption and code verification, and using secure boot and OS attestation. The Secure Networks feature leverages 5G device authentication to establish secure tunnels in accordance with pre-defined security policies. Developers can articulate security requirements through code annotations, which are then translated into tangible security mechanisms through the Security Policy Builder.

In the following, we briefly present the main tools developed in the \AISPRINT{} project that have been applied to the use cases. The organization of these tools is presented in Figure~\ref{fig:tools}. Our goal is to provide the reader with the necessary insights to understand how these tools have been integrated and used effectively, while maintaining a high-level discussion.

\subsection{\PYCOMPS{}}

\db{\PYCOMPS{}}{The PyCOMPS Distributed Computing Library (\PYCOMPS{})} is a Python library that coordinates distributed algorithms for mathematical and machine learning tasks with a simple interface. Aimed at Python developers, \PYCOMPS{} removes the complexity of parallelization, making it easy to implement large-scale machine learning workflows. At its core, it uses distributed arrays for efficient parallel operations. Designed for versatility, \PYCOMPS{} integrates applications and is compatible with various computing environments, including clouds and supercomputers, ensuring broad accessibility for developing complex machine learning solutions~\cite{dislib}. \PYCOMPS{} has been used in the Personalized Healthcare use case.

\subsection{\POPNAS}
\ifbool{double_blind}{[CustomNAS]}{Pareto-Optimal Progressive Neural Architecture Search (POPNAS)} is an advanced AutoML algorithm designed to optimize neural network architectures by balancing multiple performance objectives. It uses a progressive strategy to efficiently narrow the search space, focusing on configurations that provide the best trade-off between competing factors such as accuracy and resource usage. By aiming for Pareto optimality, \POPNAS{} effectively navigates the trade-offs in network design and identifies architectures that achieve a harmonious balance of performance metrics. This approach significantly streamlines the architecture discovery process, making it faster and more resource efficient, while ensuring that the final designs satisfy a comprehensive set of constraints. The research strategy has been considered for image classification, time series classification, and image segmentation~\cite{lomurno2021pareto, falanti2022popnasv2, falanti2023popnasv3}. \POPNAS{} has been used in the Personalized Healthcare and Farming 4.0 use cases.

\subsection{\SGDE{}}

The \ifbool{double_blind}{[GenerativeFed]}{Secure Generative Data Exchange (SGDE)} framework is designed to enhance data sharing in federated environments, such as hospitals and universities, without compromising individual privacy. Unlike traditional federated learning, which relies on the exchange of model parameters, \SGDE{} enables clients to train data generators on their local infrastructure. These generators are then uploaded to a central server, where clients can access and use them offline to create large, diverse datasets for various machine learning tasks. \SGDE{} offers key advantages over traditional federated learning. First, it provides access to representative data without sharing sensitive information. Second, it reduces the need for frequent data exchanges lowering bandwidth requirements. Finally, it improves security against data poisoning attacks by allowing easier anomaly detection in the generated data~\cite{lomurno2022sgde, lampis2023bridging}. \SGDE{} has been used in the Personalized Healthcare use case.

\subsection{\PPC{}}

The \PPC{} is designed to balance accuracy with resilience to privacy attacks in deep learning models. Specifically, it relies on adversarial training to address the dual goals of improving accuracy and reducing susceptibility to membership inference attacks by optimizing model weights. It also conducts membership inference attacks to evaluate a model's defense against such threats. This component plays a critical role in strengthening the security of models with respect to membership inference attacks on highly sensitive data. It uses AUC metrics to measure the risk of confidential information leakage during attacks and to refine training processes. The resulting models demonstrate both high effectiveness and robust defense against membership inference attacks, with test attacks approaching random guessing levels~\cite{lomurno2022utility, lomurno2023discriminative}. This tool has been used in the Personalized Healthcare use case.

\subsection{\SAIR{}}

The \SAIR{} tool manages the runtime of AI application components, focusing on optimal placement and resource selection across the computing continuum. It uses a combination of Random Search and Stochastic Local Search algorithms to adapt to runtime workload variations, aiming for cost-effective reconfigurations of initial deployments. This tool guarantees performance across multiple resources, including edge devices, cloud GPU-based virtual machines, and function-as-a-service solutions. Through real-world experiments, \SAIR{} has proven its ability to efficiently reconfigure deployments in scenarios such as wind turbine blade damage detection, providing significant cost savings over static deployments while maintaining fast execution times.
Developed in C++ with a Python interface for handling scenario-specific input and output, \SAIR{} uses performance models from the aMLLibrary. These models require a regressor file for each application component and device in the computing continuum, underscoring the tool's adaptability and precision in resource management~\cite{filippini2023space4ai}. \SAIR{} has been used in the Personalized Healthcare use case.

\subsection{\SAID{}}

The \SAID{} tool is designed to optimize the deployment of AI applications by performing design space exploration to identify the most efficient resource allocation across different computational layers. The goal is to minimize execution costs while meeting performance constraints and ensuring optimal component and resource assignments. This tool uses performance models to predict execution time for AI inference tasks and incorporates the Solid library's advanced metaheuristics, such as local search and tabu search, to refine its selection process. This tool has been used in the Personalized Healthcare use case.

\subsection{\db{\IM{}}{Infrastructure Manager}}

The \db{\IM{}}{Infrastructure Manager (IM)} is a comprehensive service for orchestrating virtual infrastructures and the applications that run on them. It manages the entire lifecycle, from resource provisioning and initial deployment to configuration, reconfiguration, and eventual termination of those resources. To meet user needs, \IM{} can integrate with catalogs such as the Virtual Machine Image Repository and Catalogue (VMRC) or the EGI AppDB to facilitate the search for the most appropriate virtual machine images (VMIs). To automate configuration tasks, \IM{} uses Ansible, leveraging a collection of roles from Ansible Galaxy. This setup ensures that the orchestration process is both efficient and tailored to specific user needs, underpinned by a high-level architecture that incorporates external dependencies for enhanced functionality. This tool has been used in the Personalized Healthcare, Maintenance and Inspection, and Farming 4.0 use cases.

\subsection{\db{\AMS{}}{AI-SPRINT Monitoring Subsystem}}

\db{\AMS{}}{The AI-SPRINT Monitoring Subsystem (AMS)} is integral to managing time series data and logs within complex, hierarchical multi-layer deployments. It ensures adherence to the \AISPRINT{} framework's constraints, offering capabilities for data collection, storage, forwarding, and analysis. Upon detecting a violation of framework constraints, \AMS{} alerts the \SAIR{} REST API endpoint. \AMS{} is crucial for monitoring key performance metrics and allows for the creation of custom metrics suited to specific needs or for consolidating data across multiple components. It enables easy incorporation of metrics into applications and facilitating the use of Python decorators for creating ad-hoc metrics and aggregating data from various sub-components. This tool has been used in the Maintenance and Inspection and Farming 4.0 use cases.

\subsection{\OSCAR{}}

\OSCAR{} is an open source platform designed to enable Functions as a Service (FaaS) for file processing applications, leveraging the scalability and flexibility of multi-cloud environments. It enables the creation of highly parallel, event-driven serverless applications that run in custom Docker containers on an elastic Kubernetes cluster. This cluster dynamically adjusts its size based on the workload, supporting efficient resource management across different deployments, including minified Kubernetes distributions such as K3s for continuous computing workflows. The platform triggers processing jobs by uploading files to MinIO, although it also supports other storage systems such as Amazon S3 and OneData. \OSCAR{}'s deployment is streamlined through the Infrastructure Manager, allowing for rapid setup and scalability. Remarkably, \OSCAR{}'s architecture is compatible with constrained devices, including Raspberry Pis, due to its support for the arm64 architecture, demonstrating its versatility and adaptability to a wide range of computing environments. \OSCAR{} has been used in the Personalized Healthcare, Maintenance and Inspection, and Farming 4.0 use cases.

\subsection{\SCONE{}}

\SCONE{} is a runtime solution designed to enhance the security of applications by integrating them into Trusted Execution Environments (TEEs) such as Intel SGX at compile time. It goes beyond simply enabling TEEs by providing transparent file system encryption and securing communications. A critical feature of \SCONE{} is its ability to perform attestation, which ensures that the application runs securely and remains unmodified within a TEE enclave. Successful attestation allows \SCONE{} to provide applications with the necessary configurations and guarantees the protection of confidential data and private keys from unauthorised access. Embedded in the executable itself, \SCONE{} also communicates with the Configuration and Attestation Service, which oversees the attestation process and manages the distribution of secrets and configurations, strengthening the security infrastructure around applications running in TEEs. \SCONE{} has been used in the Personalized Healthcare use case.

\section{\uppercase{Use Cases Analysis and Results}}
In this section, we explore the practical application of \AISPRINT{} use cases, namely Personalized Healthcare, Maintenance and Inspection, and Farming 4.0. We thoroughly examine each case, detailing the data collection methods, integration of tools, and the final evaluation stage.

\subsection{Personalized Healthcare}

\begin{figure}[t]
     \centering
     \includegraphics[width=1.\linewidth]{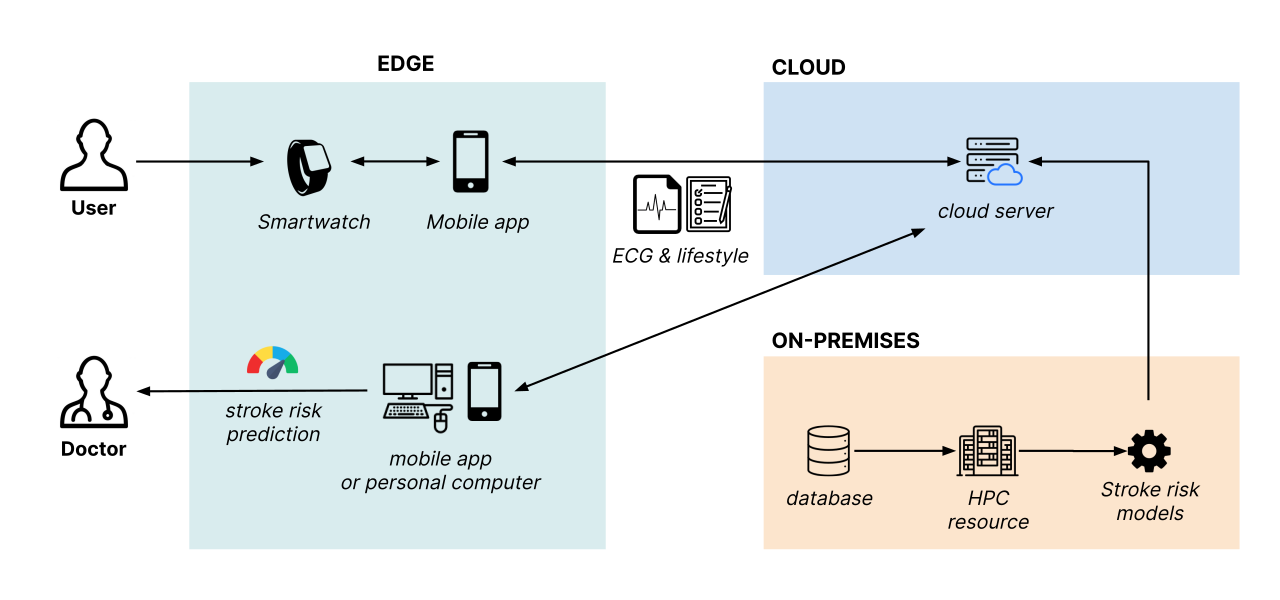}
     \caption{Architecture of the Personalized Healthcare application}
     \label{fig:healthcare}
 \end{figure}

The Personalized Healthcare use case leverages AI to improve health monitoring, with a particular focus on assessing and preventing stroke risk through the use of wearables and smartphones. This innovative method combines quantitative cardiac data with qualitative lifestyle insights to develop an AI model that improves resource utilization across edge-to-cloud platforms, ensuring efficiency and data parallelism. Recognizing the significant impact of stroke in the elderly -- a leading cause of death and disability worldwide -- the project uses non-invasive, continuous monitoring techniques to accurately distinguish atrial fibrillation signals from normal signals. It achieves this through a pilot study that demonstrates how AI can provide valuable insights for stroke care, while maintaining strict compliance with GDPR for data protection. In addition, the initiative uses federated learning to prioritize privacy and security, allowing cloud-based models to be updated with local data and synthetic private data to be shared without compromising personal information. Figure~\ref{fig:healthcare} depicts an overview of the Personalized Healthcare use case architecture.

\subsubsection{Data Collection and Processing}

The data type foundational to this use case is fully represented by the information that can be acquired from portable or wearable devices equipped with sensors. Specifically, despite the variety of vital parameters that can be monitored in real time, electrocardiogram (ECG) recordings were selected for use. The work proceeded along two parallel tracks. The first involved the collection of new data and the validation of the collection protocol, while the second focused on analyzing the best methods for processing ECG signal databases already available in the literature, with a sample size already sufficient to enable effective generalization by machine learning and deep learning models.

Data collection was carried out by anonymizing all subject information, in full compliance with the GDPR, and resulted in the recording of 15-day continuous ECGs obtained from a pilot study conducted in Spain, involving a total of 26 Spanish volunteers.
This dataset is complemented by lifestyle information collected through a questionnaire designed by \AISPRINT{}'s medical consultant at a large hospital in Spain. Co-organized by \BSC{}, a stroke foundation and a medical device manufacturer Freno Al Ictus (stroke foundation) and Nuubo (device manufacturer), two monitoring campaigns were conducted in October 2022 and March 2023. From these, a single case of dropout was observed out of the total number of data collection participants, indicating high adherence.

Among the 25 remaining individuals, 11 of the recruited subjects had a past history characterized by strokes of various types, while the remaining 14 were healthy. The age of the participants ranged from 31 to 78 years, with a mean age of 37 years. The gender distribution includes 11 females (5 with a history of stroke) and 14 males (6 with a history of stroke). The qualitative questionnaire includes a wide range of health information, facilitating the assessment of cardiovascular risk factors, lifestyle choices, and medical history. All ECG recordings underwent expert evaluation, including review by an independent diagnostic center (IDTF) and review by certified cardiac technicians (CCT).

To develop algorithms to analyze these data, the publicly available ECG dataset provided as part of the PhysioNet/CinC Challenge~\cite{goldberger2000physiobank, clifford2017af} was used.
These data, consisting of thousands of single-variate sequences from the atrial fibrillation and normal classes, were processed to handle class imbalance using shuffling-based augmentation and addressed varying sequence lengths with zero padding. Next, two different approaches led to two types of algorithm analysis. The first, remaining in the time series domain, involves feature extraction using Short-Time Fourier Transform (STFT), followed by dimensionality reduction using Principal Component Analysis, with the goal of using distributed ML models that are fast, lightweight, and benefit from strong parallelization during training. The second approach uses Fast Fourier Transform (FFT) to create spectrograms of the sequences, which are transformed to the image domain using specific hyperparameters: NFFT at 128, frequency at 2500 Hz, and overlap at 16. The resulting images are formatted to 64x64 pixels with 3 color channels, preparing for the application of 2D CNNs for image classification.

\subsubsection{Tools Integration}

This use case required the integration of several tools developed within \AISPRINT{}. For the implementation and training of machine learning models for the detection of AF, \PYCOMPS{} was extensively used both for the development of individual models and for the creation of federated learning architectures, providing an additional layer of privacy for all users.
The integration of the \POPNAS{} auto machine learning tool with the \PPC{} enabled the construction of deep neural networks specifically optimised for spectrogram image classification, providing a solution that is both highly efficient and privacy-preserving. The \SGDE{} cross-silo federated learning algorithm also demonstrated the potential for scenarios where clinics and hospitals can share synthetic data with each other. This allows individual privacy to be maintained while facilitating knowledge sharing and addressing data imbalances.
The \SAID{} and \SAIR{} evaluation tools were effectively used to assess metrics such as convergence speed, model accuracy and resource allocation. The security measures in place ensure compliance with privacy and confidentiality standards, including GDPR-compliant data collection processes and a processing environment secured by \SCONE{}.
In addition, monitoring tools and log management systems have been integrated to monitor performance, identify anomalies and support troubleshooting efforts, with successful application in the use case during the inference phase via \IM{}, \AMS{} and \OSCAR{}.

\subsubsection{Evaluation}

Overall, the Personalized Healthcare use case achieved excellent results both in terms of AI models performance and integration with the ecosystem. In particular, the dataset in time-series format has been exploited with \PYCOMPS{} to build a 1D-CNN able to achieve 90\% of accuracy with a 5-folds cross validation evaluation technique. Additionally, a 2D CNNs has been developed using the Neural Architecture Search algorithm from \POPNAS{} in combination with the \PPC{}. The result is a privacy-preserving spectrogram classifier for atrial fibrillation achieving an accuracy of 93.5\% on the test set while making privacy attacks ineffective -- only slightly better than random guessing. This results demonstrate the effectiveness of the alternative 2D time series processing. Naturally, this approach is more computationally intensive than 1D, as it deals with images rather than series, so it can easily be coupled with the first, depending on the constraints of the available infrastructure.
Concerning privacy, the \PPC{} has been used to improve model resilience against Membership Inference Attacks (MIAs)~\cite{shokri2017membership}. Specifically, this tool provides model-aware privacy guarantees on top of the infrastructure-level robustness guarantees of the \AISPRINT{} ecosystem.
This tool allows to test the effectiveness of MIAs carried out by third parties using three different attack models: threshold-based, logistic regression and random forest.
Finally, \SGDE{} has been used to generate synthetic ECG data. In fact, by utilizing synthetic data, users gain access to diverse and extensive datasets while maintaining individuals'~confidentiality. 

\subsection{Maintenance and Inspection}

\begin{figure}[t]
     \centering
        \ifbool{double_blind}{
            \includegraphics[width=1.\linewidth]{images/maintenance censored.png}
        }{
            \includegraphics[width=1.\linewidth]{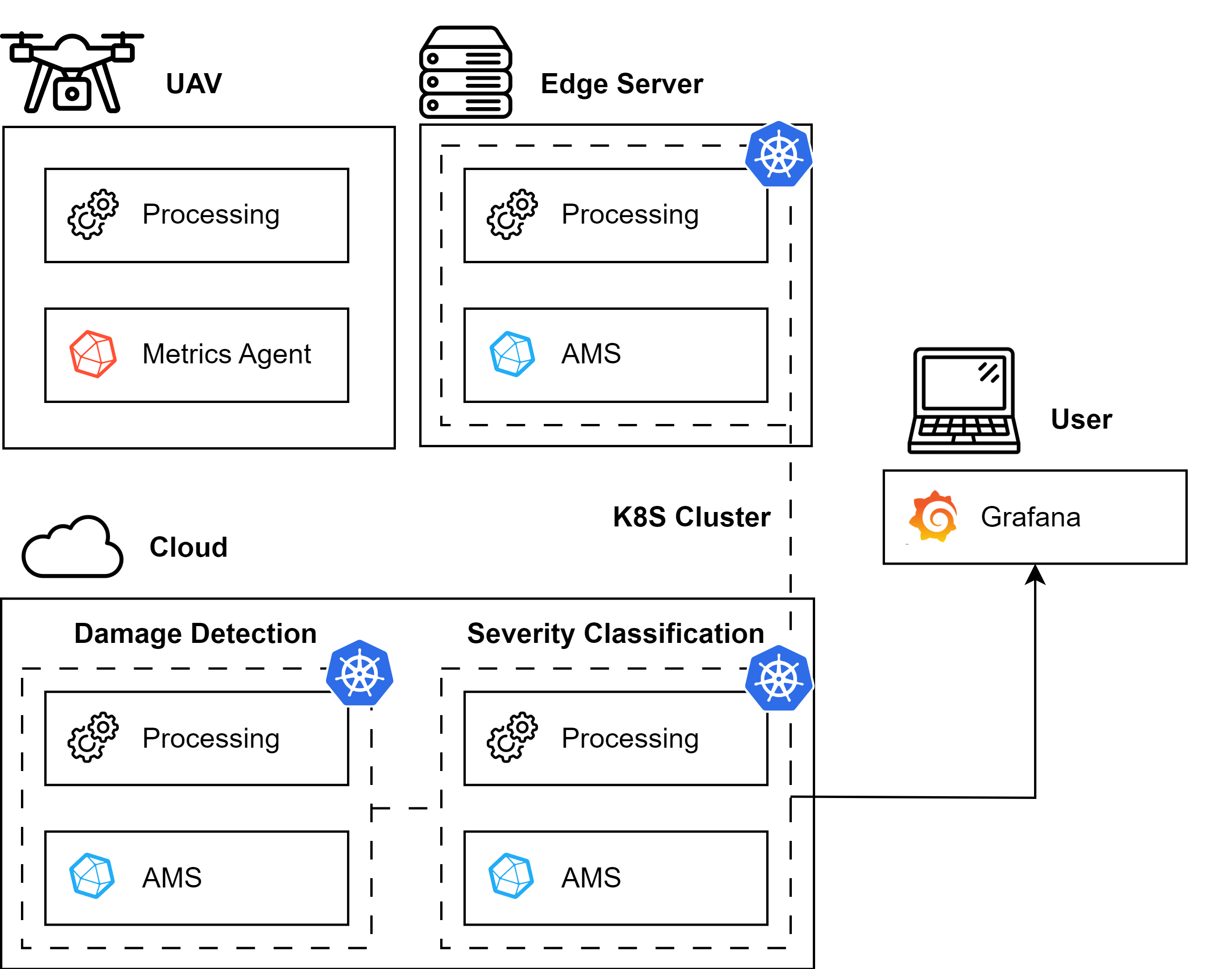}
        }
     \caption{Architecture of the Maintenance and Inspection application}
     \label{fig:maintenance}
 \end{figure}

The Maintenance and Inspection use case focuses on developing solutions in the area of infrastructure inspection through the use of proprietary AI machine learning technology. This system provides automated processes to detect, classify and monitor damage to wind turbines in real time by flying drones, with data seamlessly transferred to edge cloud architectures for comprehensive analysis. Wind turbines typically have a lifespan of 20-25 years, but damage to components such as blades, generators and gearboxes can drastically reduce this lifespan. Repair costs can range from \$300,000 for a single blade replacement to \$5 million for a complete station overhaul. To mitigate the risk of failure, thorough maintenance and inspection is critical. Currently, most inspection companies are moving toward drone-based asset inspection. The data collection process is either manual or semi-automated. Data is manually downloaded from the drone to the laptop, which is then uploaded to the cloud. AI systems play a key role by processing large amounts of data to more accurately predict and analyze maintenance needs. The \AISPRINT{} infrastructure accelerates AI-driven inspection processes by reducing operator analysis time and minimizing human error through the integration of machine learning systems, as well as the harmonious integration of cloud-based analysis and localized processing using lightweight data pattern recognition routines. Figure~\ref{fig:maintenance} shows an overview of the Maintenance and Inspection use case architecture.

\subsubsection{Data Collection and Processing}

A European branch of a US software firm developed the Maintenance and Inspection solution. As a user of \AISPRINT{}, this company leverages its own data from past projects along with its AI machine learning algorithm. The goal is to transition from cloud-based to real-time feedback, shifting data processing to drones (Unmanned Aerial Vehicle, UAVs) and edge servers. This approach uses existing legacy platform components and models, focusing on operational efficiency rather than the initial design and model training.
The standard data processing workflow for wind turbine blade inspection, supported by proprietary software, starts with gathering data. Operators capture images of the blades using drones, either manually or with semi-automated techniques. Next, operators evaluate the quality of these images. The data is then uploaded to a cloud-based tool management system. When the data becomes available, analysts begin the initial analysis with a proprietary platform to verify data completeness. Following this initial check, analysts conduct a comprehensive review of the results, making any necessary adjustments, before creating a detailed PDF report that outlines their findings.

\subsubsection{Tools Integration}
\AISPRINT{} tools enable the harmonious integration of cloud-based analytics and local processing for inspection and maintenance tasks. By leveraging the framework's capabilities, organizations can optimize efficiency by offloading complex analysis to the cloud while performing lightweight tasks locally. This approach enables fast and accurate maintenance operations, improving equipment reliability and minimizing downtime.
Offline capability allows modules (UAV, Edge, Cloud) in the field to operate in independent clusters. Model updates to edge devices are easily managed, while mission metrics such as flight times and model results are conveniently collected and analyzed by the \AMS{}. Although some UAV analytics modules fall short of the <1s response time requirement, they still operate within a 2s window between shots given the possibility of fluid UAVs. Flexible deployment mechanisms for ML models across different infrastructures are available, with the capability for significant data processing achieved through \OSCAR{}-based processing, provisioning the virtualized computing infrastructure in both private and public cloud setups using \IM{}. The system efficiently processes large amounts of data within tight timeframes, meeting the requirement to inspect a single turbine within 1-2 minutes. 
Using \OSCAR{}-based processing, the infrastructure provides scalability and flexibility between synchronous and asynchronous models. Testing on data from a single turbine inspection, with an average of 1 damage found per 3 images, confirms the system's capability. When deployed on a cluster of 3 c5.2x large AWS machines, asynchronous tasks completed processing in less than 108 seconds. Cloud computing metrics are efficiently collected and analyzed, with alerting mechanisms supported by \AMS{} and performance annotation tools. While dynamic provisioning and scaling was not tested in the use case for migrating components between the edge and the cloud or scaling up cloud resources when compute needs increase, log collection and browsing is facilitated by \AMS{}, ensuring consistent collection, centralization, and analysis of logs from all platform components.

\subsubsection{Evaluation}
The integration of \AISPRINT{} in the Maintenance and Inspection use case enhanced operations, enabling a shift from discrete, sequential, batch-oriented blocks to dynamic, near real-time processing. This facilitated a continuous flow of data, bypassing traditional bottlenecks such as asset management during transfer. The \AISPRINT{} tools provided a robust and flexible environment that was cloud provider agnostic, enabling serverless processing, efficient provisioning and infrastructure management, and continuous monitoring. This paradigm shift streamlined development and operations, improving system responsiveness and adaptability across diverse industrial inspections. 
The use case evaluation indicates significant success in the asset inspection process through the implementation of \AISPRINT{} tooling. The tests exceeded expectations, processing a significant volume of images ranging from 350 to 400 per inspection, exceeding the target of 300. The response time for quality assessment, while varying between 1.5 and 2.5 seconds, remained sufficient for smooth UAV operation and met the KPI of less than 2 seconds, albeit with slightly less accuracy. The system provided timely feedback on major damage, often within 5 minutes of inspection completion, ensuring actionable insights. It also achieved a 30\% reduction in data transmitted to the cloud, with reductions reaching 80\% for newer turbines and over 35\% for older turbines. Overall, the asset inspection process successfully met or exceeded specified KPIs, demonstrating effectiveness in image processing, response time, feedback delivery, and data transfer reduction.

\subsection{Farming 4.0}

The agricultural sector is embracing a digital revolution with the integration of GPS technologies and smart sensors paving the way for Precision Agriculture (PA). Euractiv highlights that nearly 80\% of new agricultural equipment now features PA components, representing the collaboration of around 4500 manufacturers, across 450 machine types, and involving approximately 135,000 employees worldwide. This digital transformation is projected to unlock a potential €26 billion in value from next-generation agriculture in the near future.

\begin{figure}[t]
     \centering
     \includegraphics[width=1.\linewidth]{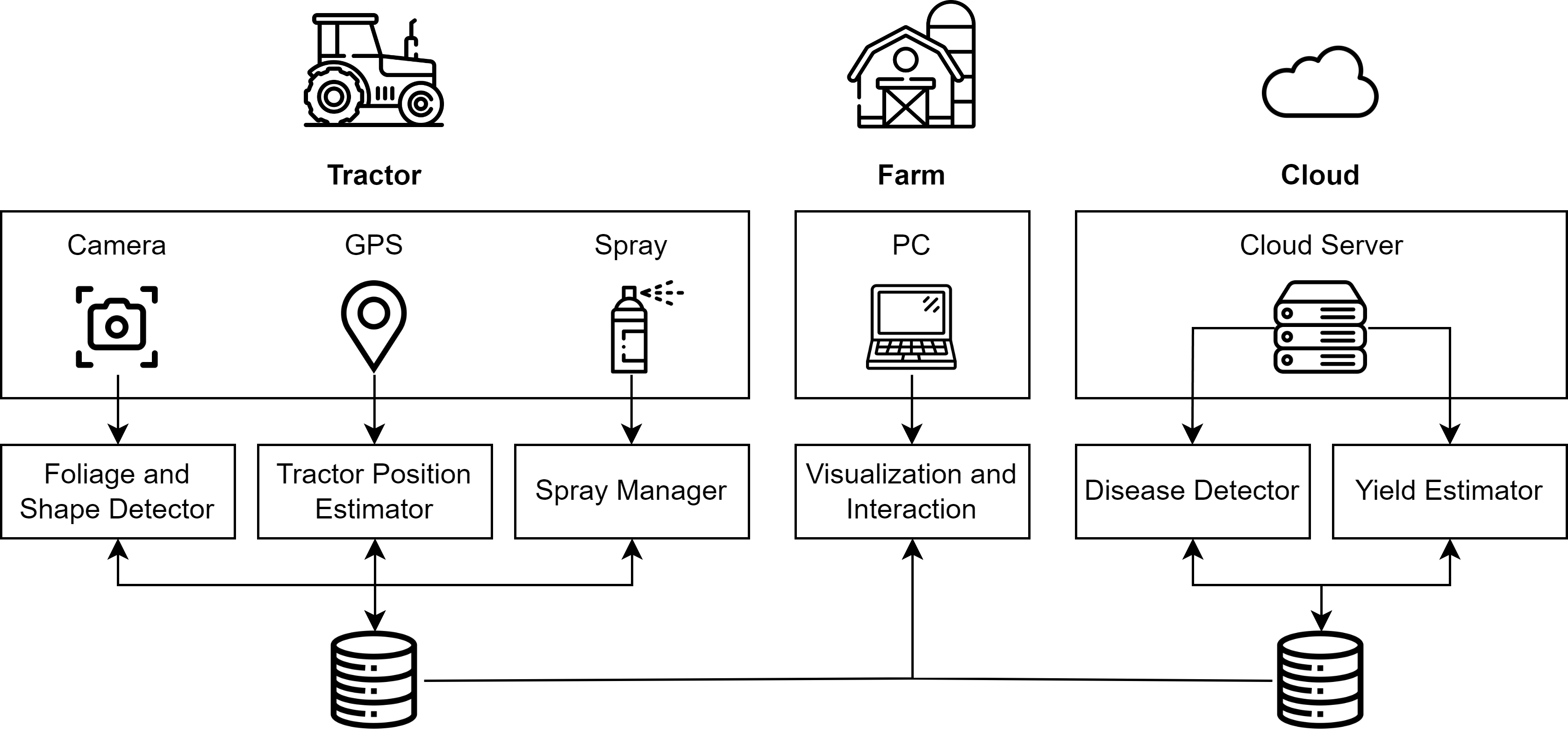}
     \caption{Architecture of the Farming 4.0 application}
     \label{fig:farming}
 \end{figure}

Within this landscape, the Farming 4.0 use case, powered by the \AISPRINT{} framework, sets out to refine phytosanitary treatments through advanced AI and edge computing technologies. The application focuses on the development and deployment of AI models that optimize the use of phytosanitary products, ensuring precise product application and minimizing environmental impact. By integrating intelligent sensors on grape harvesting machines, the use case gathers crucial data for enhancing crop monitoring and management. From these data, a streamlined image collection and analysis process has been developed between edge and cloud devices. AI models obtained via neural architecture search guarantee a state-of-the-art performance in disease detection and grape identification. Figure~\ref{fig:farming} depicts an overview of the Farming 4.0 use case architecture. 

\subsubsection{Data Collection and Processing}

The data collection for the Farming 4.0 is used to support the development and verification of an advanced spraying workflow, optimized through smart farming devices. It relies on two distinct phases. The initial phase involves a tractor equipped with smart sensors, including two Intel RealSense depth cameras (D435i), a Stereolabs ZED 2i RGB camera, a Fleet Car-PC with a GTX1050Ti GPU, a GPS Module (Reach M+ UAV RTK kit), and a Crystalfontz CFA-635 display. This setup is used to simulate the final spraying workflow's data input requirements as accurately as possible. Over the course of seven weeks in July and August 2022, this phase yielded a dataset comprising 464 RGB images, accounting for more than 7,000 labeled objects, alongside GPS, time, and acceleration data. This comprehensive data collection facilitated the detection of foliage volume and canopy shape, critical for optimizing the phytosanitary treatment workflow.

The second phase of data collection occurred in mid-2023, coinciding with the field testing of the finalized spraying workflow in a production environment. To enrich the dataset and ensure its robustness against varying environmental conditions, additional images were captured using mobile devices. This expanded dataset was labeled through a combination of automated tools and manual effort, followed by expert review, amounting to 75 images. The datasets generated during this phase are publicly accessible, providing valuable resources for ongoing research and development in precision agriculture~\cite{farmingdata1,farmingdata2}.

\subsubsection{Tools Integration}

The \AISPRINT{} suite facilitated the development of application code in an infrastructure-agnostic manner, delegating deployment to the \OSCAR{} service. This approach streamlined the packaging of the entire execution environment onto edge devices, ensuring a seamless integration process. \OSCAR{} node clusters are designed to scale automatically in response to the runtime workload, utilizing Kubernetes for efficient operations. Within the context of the farming application, classification and segmentation metrics are managed and monitored through the use of \AISPRINT{} Studio annotations and the \AMS{}. These metrics are stored into InfluxDB and are made accessible through visualization in Grafana.

For tasks such as grape disease detection and fruit volume estimation, image files gathered from sensors are automatically uploaded to the cloud, facilitating real-time data processing. The configuration of both application and device settings is streamlined through a unified YAML file, allowing for rapid modifications to be made across the system. Moreover, updates to edge devices are managed using tags, which permits targeted updates that can be specifically tailored based on criteria such as grape varieties and geographical regions, enhancing the precision and efficiency of the deployment process.

\subsubsection{Evaluation}

Without \AISPRINT{}, farming services encountered numerous challenges such as cloud-service integration, infrastructure adaptability, secure edge device deployment, and limited access to advanced AI tools. The \AISPRINT{} project has profoundly transformed the Farming 4.0 solutions by effectively tacking these issues. 
Notably, key achievements include the efficient deployment of smart farming devices on tractors leveraging edge computation and cloud infrastructure. The system successfully managed performance metrics collection and data transfer with a processing time of 0.8 seconds per image while handling over five images in parallel. Additionally, data processing has been carried on multiple cloud platforms to showcase the framework flexibility with rapid transition. The best foliage segmentation model was developed using the neural architecture search algorithm from \POPNAS{}, obtaining a mean accuracy of 96.8\% and mean intersection-over-union of 94.2\%.

\section{\uppercase{Lessons Learned and Recommendations}}
The \AISPRINT{} initiative, through its application in Personalized Healthcare, Maintenance and Inspection, and Farming 4.0, has clearly demonstrated the ability of AI to revolutionize various industries by improving efficiency, adaptability, and performance. In the Personalized Healthcare use case, the use of \AISPRINT{} enabled significant advances in stroke care, where its precision in detecting atrial fibrillation from ECG signals with high accuracy. This success was underpinned by the integration of \PYCOMPS{} for distributed machine learning and the innovative use of federated learning, which together improved predictive performance while ensuring data security through encrypted transmissions. The use of \OSCAR{} for on-demand inference and the strategic implementation of quality annotations early in the design process were critical to meeting the specific requirements of the healthcare sector.
As for the Maintenance and Inspection use case, the project encountered system orchestration challenges due to the resource constraints of components such as the Nvidia Jetson Nano. This led to a shift towards the development of a lightweight, dedicated component, highlighting the importance of flexibility and the need to tailor solutions to specific operational contexts. This adaptation underscores a broader lesson within the \AISPRINT{} journey: the need for pragmatic and resource-conscious approaches to technology deployment.
In agriculture, the Farming 4.0 use case demonstrated the transformative potential of \AISPRINT{} in smart farming devices, where it significantly reduced the deployment effort and enhanced the real-time processing capabilities of farming equipment. The use of \OSCAR{} and the \POPNAS{} tool for AI model optimization exemplifies the project's success in combining rapid data processing with actionable insights for on-the-fly decision making, thereby optimizing agricultural output and resource use.
Together, these implementations validate \AISPRINT{}'s role as a catalyst for innovation across industries, providing lessons on the value of distributed machine learning, the need for secure and efficient data processing, and the benefits of adaptive, context-aware technology solutions. Through its diverse applications, \AISPRINT{} not only addresses specific industry challenges, but also sets a precedent for future AI-driven efforts, emphasizing a holistic approach that combines technical excellence with operational pragmatism.

\section{\uppercase{Conclusions}}
\label{sec:conclusion}
The \AISPRINT{} framework has demonstrated its versatility and effectiveness in a variety of industry scenarios, such as Personalized Healthcare, Maintenance and Inspection, and Agriculture 4.0. This adaptability is underscored by significant achievements such as the development of a GDPR-compliant stroke risk assessment system in healthcare, real-time windmill blade damage detection for maintenance, and improved pesticide application efficiency in agriculture, all facilitated by AI-enabled tools.
In Personalized Healthcare, the framework demonstrated its ability to not only adhere to strict privacy standards, but also deliver high processing efficiency, illustrating the power of its applications in critical healthcare areas. In Maintenance and Inspection, there was a notable shift to a near real-time operational model, making development and operations more seamless with the help of the \AISPRINT{} toolkit. In Farming 4.0, the project achieved its goal of increasing the efficiency of pesticide spraying by deploying smart farming devices that have yielded positive results.
These achievements underscore the significant contributions of the \AISPRINT{} project to the field of artificial intelligence, addressing unique challenges across different sectors. The experience gained from these successes, as well as the challenges and lessons learned, provide valuable insights for future AI developments. This body of work solidifies the \AISPRINT{} project's impact on the ever-evolving artificial intelligence landscape.

\bibliographystyle{apalike}
\bibliography{example}

\end{document}